\documentstyle[11pt,epsf]{article}

\begin{document}

\baselineskip 0.8cm

\title{ \Large\bf
 Analyzing Chiral Symmetry Breaking \\
 in Supersymmetric Gauge Theories
}
\author{
 Thomas Appelquist\thanks{E-mail: twa@genesis3.physics.yale.edu \,.},
 Andreas Nyffeler\thanks{E-mail: nyffeler@genesis2.physics.yale.edu ;
 current address: DESY-IfH Zeuthen, Platanenallee 6, D-15738 Zeuthen,
 Germany.},
 and Stephen B. Selipsky\thanks{E-mail: selipsky@jdub.wustl.edu ;
 current address: Dept.\ of Physics, Washington University,
 1 Brookings Dr., Campus Box 1105, Saint Louis, MO 63130-4899, USA.}
\\ \\ {\it
 Dept.\ of Physics, Yale University, New Haven, CT 06520-8120, USA}
}

\date{November 7, 1997}

\maketitle
\begin{picture}(0,0)(0,0)
\put(365,245){hep-th/9709177}
\put(365,230){YCTP--P12--97}
\put(365,215){WUSTL--HEP--97--55}
\end{picture}
\vspace{-24pt}

\begin{abstract}
   We compare gap equation predictions for the spontaneous breaking of
global symmetries in supersymmetric Yang-Mills theory to nonperturbative
results from holomorphic effective action techniques.
In the theory without matter fields, both approaches describe the formation
of a gluino condensate.  With $N_f$ flavors of quark and squark fields,
and with $N_f$ below a certain critical value, the coupled gap equations
have a solution for quark and gluino condensate formation,
corresponding to breaking of global symmetries and of supersymmetry.
This appears to disagree with the newer nonperturbative techniques,
but the reliability of gap equations in this context and whether the
solution represents the ground state remain unclear.
\end{abstract}

\section{Introduction}

   Spontaneous breaking of global symmetries in a gauge field theory sets in
only when the gauge forces become strong.  Accordingly, nonperturbative
methods must be brought to bear, and the problem remains only partially
understood.  Approaches include lattice methods, semiclassical methods,
and $1/N$ expansions.
In addition, a simple approach based in Feynman graphs has been used with
apparent success for many years.  It comes in several closely related forms,
including Schwinger--Dyson or gap equations, and ``most attractive channel''
analyses.  These share the common idea that the forces responsible for
symmetry breaking can be computed in a loop expansion.  The lowest order
contribution is typically one-particle exchange, with higher order
contributions given by two particle irreducible graphs \cite{CJT,Peskin1}.
This can be applied to a most attractive channel force analysis or
equivalently the kernel of a gap equation, or the CJT effective potential
\cite{CJT} for composite operators.

   Because the coupling must be relatively strong, there is no obvious small
parameter in the expansion and this approach has been met with justifiable
skepticism.  Some evidence in favor of its utility, however, is provided by
estimates of the second order contribution to the kernel \cite{ALM,Nash}.
When the coupling is just strong enough to trigger the breaking, second order
corrections are relatively small (less than 20\%).  With this check in mind,
and in the absence of clearly superior alternatives, gap equations have been
widely used in recent years in the study of dynamical electroweak symmetry
breaking \cite{MexicoBrazil}.  Still, their validity remains open to question.

   Recent advances \cite{SeibWitReview} in nonperturbative methods for
supersymmetric gauge theories provide an independent framework in which
to test the validity of the gap equation approach.
The new methods lead to conclusions about spontaneous breaking of global
symmetries in supersymmetric theories, which can be compared to gap equation
results.  This letter reports the results of such a comparison, first for
the case of ${\cal N} = 1$ supersymmetric Yang-Mills $SU(N_c)$ theory with
no matter fields (pure SQCD), and then for ${\cal N} = 1$ SQCD including
matter fields.

\section{Pure Supersymmetric QCD}

   The Lagrangian for pure SQCD contains only gluon and gluino degrees
of freedom (in Wess-Zumino gauge after eliminating auxiliary fields):
\begin{equation} \label{SYM-L}
 {\cal L}_{SYM} \ =\ -{1\over 4}\, G_{\mu\nu}^a G^{\mu\nu \, a}
 \ +\ {\theta g^2 \over 32 \pi^2}\, G_{\mu\nu}^a \widetilde G^{\mu\nu \, a}
 \ +\ {1\over 2}\, \bar\lambda^a i \slash \!\!\!\! D \lambda^a
 \ +\ {\rm gauge\ fixing\ +\ ghosts\ ,}
\end{equation}
where the canonically normalized Majorana spinor $\lambda$ describes the
gluino field, $\theta$ is the vacuum angle for the nonabelian gauge
field $G_{\mu\nu}$, and
 $iD_\mu \lambda = i\partial_\mu \lambda - g[A_\mu , \lambda]\,$.
This theory has an anomalous $U(1)_R$ global symmetry,
$\lambda \rightarrow \exp{(i\vartheta)\lambda}$,
broken by instanton effects to $Z_{2 N_c}$.

   Various studies indicate the formation of a gluino condensate
\cite{gluinocond,FinnelPouliot},
which spontaneously breaks\footnote{\
 Note however a suggestion to the contrary \cite{nogluino}.}
the discrete $Z_{2 N_c}$ symmetry to $Z_2$.
Nonperturbative holomorphic calculations derive this result starting from
an exact Wilsonian effective superpotential for $N_f = N_c - 1$, obtained
\cite{FinnelPouliot,Cordes,Affleck-Dine-Seiberg} from supersymmetric and
global symmetry constraints, assuming the existence of a nonperturbative,
supersymmetric regulator.
Successively integrating out massive flavors \cite{integrout} results in an
effective superpotential for $N_f = 0$.  Differentiating with respect to the
supersymmetric coupling $\tau \equiv\, i 4\pi/g_h^2 +\, \theta/2\pi$ then
provides an exact formula \cite{FinnelPouliot} for the gluino condensate,
 $\langle \lambda^a_h \lambda^a_h \rangle = 32 \pi^2 \Lambda_h^3$.  Here,
the subscript $h$ denotes a ``holomorphic'' normalization \cite{gjacobian}
($1/g_h^2$ on the kinetic term) for a Weyl spinor, and
 $\Lambda_h \equiv M \exp{[2\pi i \tau / (3 N_c)]}$
is defined in terms of a holomorphic gauge coupling $g_h^2$ defined at an
ultraviolet scale $M$ in the dimensional reduction scheme \cite{FinnelPouliot}.
For a canonical field normalization, rescaling $\lambda_h = g(M^2)\, \lambda$
requires the canonical coupling $g$ to satisfy \cite{newg}
 $8\pi^2/g_h^2(M^2) = 8\pi^2/g^2(M^2) + N_c \ln{g^2(M^2)}$,
due to the transformation's anomalous Jacobian \cite{gjacobian}.
We can accordingly express the holomorphic result as
\begin{equation} \label{gluinoVEV}
 {1\over 2}\ \langle \bar\lambda^a \lambda^a \rangle\ =\
 {32\pi^2 \over g^2(M^2)}\, \Lambda_h^3 \ =\
 {32\pi^2 \over g^4(M^2)}\, \Lambda_1^3
\end{equation}
(with canonically normalized, Majorana spinors).
The one-loop $\beta$ function solution for $g$ diverges at the ``confinement''
scale $\Lambda_1 \equiv M \exp{[-8\pi^2 / (3 N_c\, g^2(M^2))]}$.

   Gap equation techniques, too, indicate a $Z_{2 N_c}$-breaking condensate,
signalled by a dynamical gluino mass.  We write the gluino inverse
propagator as
 $A_\lambda(p^2)\, [p\!\!\!\slash - \widetilde\Sigma_\lambda(p^2)]\, i^{-1}$,
with the wave function factor $A_\lambda(p^2)$ defined for renormalized fields
so that $A_\lambda(\mu^2) = 1$, and with the dynamical mass
$\Sigma_\lambda(p^2) = A_\lambda(p^2)\, \widetilde\Sigma_\lambda(p^2)$.
In a gauge with gluon propagator
 $-i [g^{\mu\nu} + (\xi - 1)k^\mu k^\nu /k^2]/k^2$,
the gap equation in Euclidean space takes the form
\begin{equation} \label{gluinogap}
 \widetilde\Sigma_\lambda(p^2) \ =\ C_2^{\rm adj}\, g^2\, (3 + \xi)
 \int{{d k^2\over 16 \pi^2} \,
 {k^2 \over {\cal M}^2}\, {A_\lambda ({\cal M}^2)\over A_\lambda(m^2)} \,
 {\widetilde\Sigma_\lambda(k^2) \over
    k^2 + \widetilde\Sigma^2_\lambda(k^2)}}\ ,
\end{equation}
where ${\cal M}$ and $m$ are respectively the larger and smaller of $k$ and
$p$.  (The angular integrations have been performed in four dimensions,
after approximating $A_\lambda([k-p]^2)$ by $A_\lambda({\cal M}^2)$,
and inserting an extra factor of $A_\lambda({\cal M}^2)$ in anticipation of
the linearization discussed below.)  The structure of this equation is similar
to that for the quark in ordinary QCD \cite{MexicoBrazil,ALM}, differing only
in that the group factor $C_2^{\rm adj} = N_c$ replaces $C_2^{\rm fund} =
 (N_c^2 - 1)/(2 N_c)$.  We have neglected running of the coupling and of
the gauge parameter \cite{MexicoBrazil}, a crude approximation adequate to
establish the existence of a symmetry-breaking solution in this theory and
to set the stage for the presence of matter fields.
Here, the approximation requires an ultraviolet cutoff on the integral.
When matter fields are included, an infrared fixed point will govern the
transition and justify the neglect of running.

   A nontrivial solution to Eq.~(\ref{gluinogap}) requires the coupling to
exceed a critical value, near which the dynamical mass vanishes continuously
\cite{ATW96}.  This allows the critical coupling to be determined by analyzing
the equation in the regime $p \gg \widetilde\Sigma_\lambda(p^2)$, where loop
momenta $k \gg \widetilde\Sigma_\lambda(k^2)$ dominate the integral and
linearization in $\widetilde\Sigma_\lambda$ is a good approximation.
We can then also neglect corrections to the massless renormalization group
formula $A_\lambda(p^2) \approx (\mu^2/p^2)^{\gamma_\lambda}$,
where the scaling exponent equals the gluino-field anomalous dimension
\begin{equation} \label{gluino-gamma}
 \gamma_\lambda \ =\ C_2^{\rm adj}\, \xi \, {g^2 \over 16 \pi^2}\ .
\end{equation}
We find solutions by \cite{ALM} inserting a scaling form
 $\widetilde\Sigma_\lambda(p^2) \approx
 \widetilde\Sigma_\lambda(\mu^2)\, (\mu^2/p^2)^{b_\lambda}$.
The scaling exponent $b_\lambda$, in leading approximation just (half)
the gluino-mass anomalous dimension, obeys a $\xi$-independent equation
to first order in $g^2$:
\begin{equation} \label{gluino-b}
 b_\lambda\, (1 - b_\lambda) \ =\ 3\, C_2^{\rm adj}\, {g^2 \over 16\pi^2}
 \ +\ {\cal O}(g^4) \ .
\end{equation}
A solution requires the right hand side of this equation to be large enough
for $b_\lambda$ to become complex \cite{MexicoBrazil}, corresponding to the
mass anomalous dimension reaching unity.  The discrete global symmetry is
predicted to break if the coupling exceeds the resulting critical minimum
value,
 $\alpha_{\rm cr} \equiv g_{\rm cr}^2/4\pi
 = \pi/(3 C_2^{\rm adj}) = \pi/(3 N_c)$.
Higher order corrections to the kernel of the gap equation may be
small enough not to affect this conclusion qualitatively.
In nonsupersymmetric QCD, the next order corrections are less than 20\%
\cite{ALM,Nash}.

   In reality, the coupling runs; this effect can be incorporated in a
WKB-like approximation \cite{MexicoBrazil} which predicts nonzero
solutions at scales where the coupling exceeds the above critical value.
Since the theory is asymptotically free and presumably confining, that
will occur at momenta of order the confinement scale.\footnote{\
 The Wilsonian effective coupling has been argued to be controlled
 by an exact $\beta$ function \cite{NSVZ} for $N_f = 0$:
 $d g / d\ln{\mu} = -(3 N_c / 16\pi^2)\, g^3 / (1 - N_c g^2/(8\pi^2) )$.
 Well before reaching the singularity in this $\beta$ function, the
 running coupling will reach the above $g_{\rm cr}$.  The $\beta$ function's
 expansion parameter there is $N_c \alpha_{\rm cr} / (2\pi) = 1/6$.}
The symmetry is thus predicted to break, and the confinement scale sets
the size of $\widetilde\Sigma_\lambda$ via the full nonlinear gap equation.
Inserting this result into the operator product formula for the gluino
condensate leads to the estimate
\begin{equation} \label{OPcondensate}
 \langle \bar\lambda^a \lambda^a \rangle\ \approx\
 4\, (N_c^2 - 1) \int{ d^4 k \over (2\pi)^4 }\,
 {\widetilde\Sigma_\lambda(k^2)\over k^2 + \widetilde\Sigma_\lambda^2(k^2)}\,
 {1\over A_\lambda(k^2)}\ \sim\ (N_c^2 - 1)\, {\Lambda_1^3 \over 4\pi^2}\ .
\end{equation}
The last rough estimate, neglecting logarithmic factors, arises from
dimensional analysis with a nonzero solution to the gap equation,
identifying the confinement scale with $\Lambda_1$ in Eq.~(\ref{gluinoVEV}).
For large $N_c$, we may write this estimate in the form
\begin{equation} \label{Ncondensate}
 \langle \bar\lambda^a \lambda^a \rangle\ \sim\ {1\over 2}
 \left( {N_c g^2(M^2)\over 8\pi^2} \right)^2\,
 {32 \pi^2 \over g^4(M^2)}\, \Lambda_1^3 \ ,
\end{equation}
which has the same $N_c$ scaling (with fixed $N_c g^2(M^2)$) as
Eq.~(\ref{gluinoVEV}).

   We note that a complete gap-equation prediction of symmetry breaking would
require comparing the broken solution to the $\widetilde\Sigma_\lambda = 0$
solution.  Establishing the former to be the ground state would require
settling currently unresolved issues of gauge dependence, resulting from
truncating the effective action or the infinite set of coupled
Schwinger-Dyson equations.
(Even in QCD, gap equation techniques face the same question.)
Assuming the broken solution to be the ground state (with zero energy for
unbroken supersymmetry), the gap equation and holomorphic approaches both
predict that the discrete $Z_{2 N_c}$ symmetry spontaneously breaks to
$Z_2$ via a gluino condensate.

\section{Supersymmetric QCD with $N_f$ Flavors}

   Including quark supermultiplets with $N_f$ flavors provides more scope
for comparing the two approaches.  Again using component notation with
auxiliary fields eliminated, we add $N_f$ quark flavors to the Lagrangian,
as Dirac spinors in the $SU(N_c)$ fundamental representation; their left-
and right-handed components are respectively associated with scalar
superpartners, $\phi$ and $\widetilde\phi^*$:
\begin{eqnarray} \label{SQCD-L}
 {\cal L}_{SQCD} \ =\ {\cal L}_{SYM} \,
 +\, \bar\psi (i \slash \!\!\!\! D - m_0) \psi\,
 +\, |D_\mu \phi|^2 \ +\ |\widetilde D_\mu \widetilde\phi|^2 \,
 -\, m_0^2 \left(\phi^* \phi + \widetilde\phi \widetilde\phi^* \right)
\nonumber\\
 -\, {g^2\over 2}
 \left( \phi^* T^a \phi - \widetilde\phi T^a \widetilde\phi^* \right)^2
 \, -\, ig\sqrt{2} \left( \phi^* T^a \bar\lambda^a P_L \psi +
     \widetilde\phi T^a \bar\lambda^a P_R \psi - {\rm h.c.} \right)\ ,
\end{eqnarray}
where $i D_\mu \equiv i\partial_\mu - g A_\mu^a T^a$.  In the limit of
vanishing bare masses $m_0 \rightarrow 0$, the global symmetries are:
a discrete parity symmetry,
 $(\psi_L , \phi) \leftrightarrow (\psi_R , \widetilde\phi^*)$;
chiral $SU(N_f) \times SU(N_f)$ with quarks and squarks rotated together;
baryon number $U(1)_B$ for quarks and squarks;
and the anomaly-free subgroup $U(1)_X$
of the gluino/squark $U(1)_R$ and the quark/squark axial $U(1)_A$.
Although these symmetries forbid perturbative mass generation,
nonperturbative condensates could break the symmetries, generating fermion
and scalar dynamical masses and mixing the left- and right- scalars.
The scalar mixing would also require nonvanishing dynamical gluino mass,
since $U(1)_X$ combines gluino and axial rotations.

   An important step in the nonperturbative study of supersymmetric theories
was Witten's observation \cite{WittenIndex} that in a massive theory the
difference between the number of bosonic and fermionic zero-energy states
is conserved\footnote{\
 For the exactly massless theory, zero modes can come in from infinity
 and leave the index ill-defined.  This is avoidable \cite{indexmassless}
 if $N_f$ is an integer multiple of $N_c$.}
(the Witten index), with a nonzero value ($N_c$ for massive SQCD) implying
unbroken supersymmetry.  In that case, supersymmetric Ward identities for
$N_f > 0$ leave fermion condensates proportional to the bare mass.
Specifically, for the gluino condensate the Konishi anomaly \cite{Konishi}
gives ${1\over 2}\, \langle \bar\lambda \lambda \rangle = (g^2 / 16 \pi^2)\,
 m_0\, \langle \phi \widetilde \phi \rangle$.
For the quark condensate \cite{SUSYWard}, each flavor $i$ satisfies
$\langle \bar\psi_i \psi_i \rangle = m_0\, \langle
 \phi_i^* \phi_i \, +\, \widetilde\phi_i \widetilde\phi_i^* \rangle$.
The massless theory thus forbids fermion condensates, assuming a well-defined,
nonvanishing index at $m_0 = 0$, and assuming squark bilinear expectations
are not too singular when $m_0 \rightarrow 0$.

   The holomorphic effective action approach \cite{SeibWitReview}
provides a more complete description of SQCD with $N_f$ flavors.
Assuming the existence of a nonperturbative, supersymmetric regulator,
it leads to a self-consistent picture for various values of $N_f$.
Many physically distinct but (if $N_f \ge N_c$) degenerate vacua
form a moduli space, on which generically the chiral symmetry breaks,
but at special points (when expectation values vanish) is preserved.
\\ \noindent
$\bullet$ For $N_f > 3 N_c$, there is no asymptotic freedom, and the quarks,
squarks, gluons and gluinos become noninteracting at large distances.
\\ \noindent
$\bullet$ For ${3\over 2} N_c < N_f < 3 N_c$, the theory is asymptotically
free and the coupling runs to a ``superconformal'' infrared fixed point.
There is no confinement in this ``nonabelian Coulomb phase'', and the global
symmetries remain unbroken at the origin of moduli space (where the squark
vacuum expectation values vanish).  The spectrum is described by interacting,
massless quarks, squarks, gluons, and gluinos.
\\ \noindent
$\bullet$ For $N_c + 2 \le N_f \le {3\over 2} N_c$, there is no confinement,
but the gauge theory is strongly interacting in the infrared.  The spectrum
is best described in terms of massless, composite mesons and baryons of a
local, infrared-free, dual gauge theory.  At the origin of moduli space
the theory leaves unbroken all the original global symmetries.
\\ \noindent
$\bullet$ For $N_f = N_c$ or $N_f = N_c + 1$, the theory confines and the
spectrum consists of massless composite particles corresponding to meson
and baryon fields composed of the original matter fields.  For $N_f = N_c$,
either $U(1)_B$ or chiral $SU(N_f) \times SU(N_f)$ must break.  For
$N_f = N_c + 1$, there can be confinement without chiral symmetry breaking.
\\ \noindent
$\bullet$ For $1 \le N_f < N_c$, a nonzero superpotential
 \cite{Affleck-Dine-Seiberg}
lifts the degeneracy and fixes the ground state at arbitrarily large
scalar expectation values (for $m_0 \rightarrow 0$).
\\ \noindent
An important feature of these results is that for $N_f > N_c$,
the global symmetries associated with massless quarks,
squarks and gluinos can all remain unbroken.

   We next use the gap equation framework to study the SQCD theory with matter.
A previous investigation \cite{shamir} concentrated on the quark and squark
dynamical masses, neglecting the gluino condensate by assuming the matter
fields to occupy a high dimensional representation.
Here we will retain all condensates, to obtain a set of coupled gap equations
for the quarks, gluinos, and squarks.  The most attractive interactions
occur in channels that preserve local $SU(N_c)$ and parity,
so we will not consider color-breaking condensates, or differing dynamical
masses for the scalars $\phi$ and $\widetilde\phi$ (Eq.~\ref{SQCD-L}).
We will also exclude nonzero vacuum values for the squark fields.
Supersymmetry requires equal bare masses for the quarks and squarks, which
we take to vanish; but the dynamical masses can differ, since the component
notation does not maintain manifest supersymmetry in off-shell Green
functions.

   For small enough coupling, the running will be determined by the
(scheme-independent) two-loop $\beta$ function \cite{SQCDbeta}
\begin{equation} \label{susybeta}
 { d g \over d\ln{\mu} }\ =\ -{g^3 \over 16\pi^2}\, (3 N_c - N_f)\ -\
 {2 g^5\over (16 \pi^2)^2} \bigl( 3 N_c^2  - 2 N_c N_f + N_f/N_c \bigr)\
 +\ \cdots\ ,
\end{equation}
which displays an infrared fixed point when
 $3 N_c/(2 - N_c^{-2}) < N_f < 3 N_c$.  It occurs at
\begin{equation}\label{fixedpoint}
 \alpha_* \ =\ 2\pi\, { 3 N_c - N_f \over 2 N_c N_f - 3 N_c^2 - N_f/N_c }\ ,
\end{equation}
a small value for $N_f$ sufficiently close to $3 N_c$.
The coupling approaches this fixed point at scales below some intrinsic
scale $\Lambda$ governing the solution to Eq.~(\ref{susybeta}).
If the fixed point coupling exceeds a critical value,
then the gap equations will exhibit nontrivial solutions.
Since the dynamical mass vanishes continuously there,\footnote{\
 Ref.~\cite{ATW96} argued that although the dynamical mass vanishes
 continuously, there are no light degrees of freedom in the symmetric phase
 and therefore the transition is not, strictly speaking, of second order.}
the transition will be governed by the fixed point, and the
running of the coupling may be neglected to first approximation.
Furthermore, the gap equations may be linearized in the neighborhood of
the transition.

\begin{figure}[hbt]
\hfil{
 \epsfysize = 2in
 \epsfbox{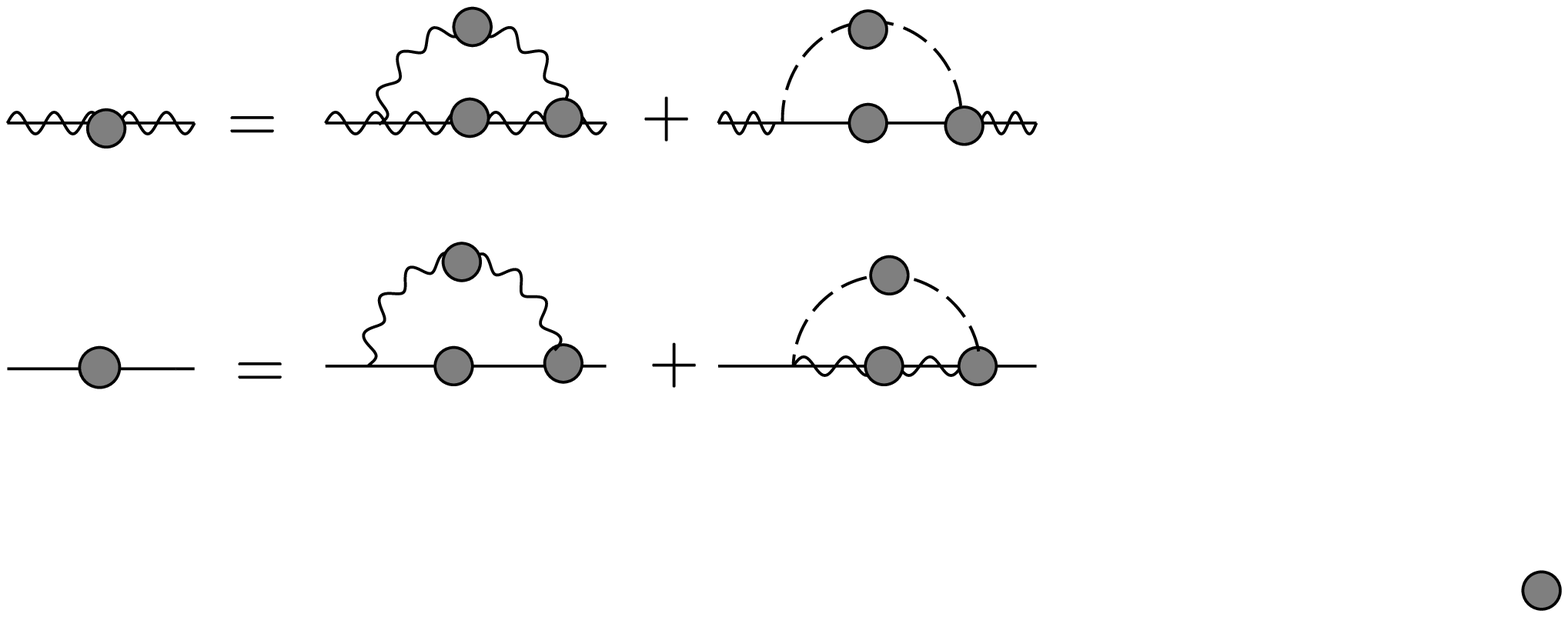}
 }\hfil
\caption{
 Gap equations for the gluino and for the quark masses.
 Solid lines are quarks, dashed lines are squarks, wiggled
 lines are gluons, and wiggled-on-solid lines are gluinos.
 Blobs symbolize the full propagators and vertex structures.
 The scalar exchange graphs give integrands proportional to higher
 powers of dynamical mass, and are dropped in the linearized
 approximation.
} \label{fig1}
\end{figure}

   The gluino and quark gap equations are shown in Fig.~1.  To lowest order
in the kernel and in the linearized approximation relevant to the study of
the transition and the determination of the critical coupling, they each
receive a contribution only from the first graph (gluon emission and
reabsorption).  The gluino gap equation is simply the linearized version of
Eq.~(\ref{gluinogap}), and the corresponding quark equation takes the form
\cite{ALM}
\begin{equation} \label{quarkgap}
 \widetilde\Sigma_\psi(p^2) \ =\ g^2\, C_2^{\rm fund}\, (3 + \xi)
 \int {d k^2 \over 16\pi^2} \,
 {k^2 \over {\cal M}^2}\, {A_\psi ({\cal M}^2)\over A_\psi (m^2)} \,
 { \widetilde\Sigma_\psi(k^2) \over k^2 }\ ,
\end{equation}
where $g^2$ is now the value of the coupling at the infrared fixed point.
For both, the intrinsic scale $\Lambda$ is effectively an ultraviolet cutoff
in this approximation: the functions $\widetilde\Sigma(p^2)$ steepen there
from near-critical behavior $\sim 1/p$, to asymptotic behavior $\sim 1/p^2$.
The gauge parameter is also in general a function of momentum, governed by
a renormalization group equation.  When $g^2$ approaches its infrared fixed
point, the gauge parameter has one as well.  We denote this fixed-point value
simply by $\xi$, and in this approximation take it outside the integral.
The wave function factors $A(p^2)$ are computed in the massless theory,
dropping corrections subleading in mass.  They are defined to be unity at
the renormalization scale $\mu^2$.

   For the scalars, we obtain a gap equation for the diagonal mass
$\widetilde\Sigma_\phi^2(p^2)$,
appearing in both $\langle \phi(p)\, \phi^*(-p)\rangle$
and $\langle\widetilde\phi^*(p)\, \widetilde\phi(-p) \rangle$:
\begin{eqnarray}\label{squarkDgap}
 && p^2 \Bigl( A_\phi(p^2) - Z_\phi \Bigr)\,
 +\, A_\phi(p^2)\, \widetilde\Sigma_\phi^2(p^2)\
\\
 && =\ C_2^{\rm fund} \int { d k^2 \over 16 \pi^2 }\, g^2
 \Biggl[
  { A_\phi^2({\cal M}^2) \over A_\phi(k^2) }
  \Bigl\{ 1 - \xi + (\xi - 3) {m^2\over {\cal M}^2} \Bigr\}
  { k^2 \over k^2 + \widetilde\Sigma_\phi^2(k^2) }\,
 +\, (3 + \xi)\, A_\phi({\cal M}^2)\, { A_g({\cal M}^2) \over A_g(k^2) }
 \Biggr.
\nonumber\\
 && \Biggl. \qquad\quad -\,
  { A_\phi({\cal M}^2) A_\psi({\cal M}^2) \over A_\psi(k^2) }
  { 2 \over k^2 + \widetilde\Sigma_\psi^2(k^2) }
 \Bigl\{ k^2 + {m^2\over p^2} \, (k^2 - p^2)\, -\,
  \widetilde\Sigma_\lambda^2({\cal M}^2) \cdot [1+{\rm sign}\, (k^2 - p^2)]
 \Bigr\}
 \Biggr] \nonumber
\end{eqnarray}
where the wave function renormalization constant $Z_\phi$ implements
$A_\phi(\mu^2) = 1$, and $A_g(k^2)$ represents the gluon wave function
factor.\footnote{\
 Here, $g$ and $\xi$ are in general functions of ${\cal M}^2$.  As in the
 fermion equations, this dependence can be neglected if the integrals are
 dominated by an infrared fixed point.}
The dynamical mixing mass $\widetilde\Sigma_X^2(p^2)$
in $\langle \widetilde\phi(p) \phi(-p) \rangle$ satisfies
\begin{eqnarray}\label{squarkXgap}
 && \widetilde\Sigma_X^2(p^2) \ =\
 C_2^{\rm fund} \int {d k^2 \over 16 \pi^2}\,
 { g^2 \over {\cal M}^2 }
 \Biggl[
  { A_\phi({\cal M}^2) \over A_\phi(m^2) }\,
  \widetilde\Sigma_X^2(k^2)\,
  \Bigl\{ (\xi + 1) {\cal M}^2 + (3 - \xi) m^2 \Bigr\}
  { k^2 \over \left( k^2 + \widetilde\Sigma_\phi^2(k^2) \right)^2 }
 \Biggr.
\nonumber\\
 && \Biggl. \hskip 9em +\, 4\,
 { A_\phi({\cal M}^2) A_\psi({\cal M}^2) \over A_\phi(p^2) A_\psi(k^2) }\,
 \widetilde\Sigma_\psi(k^2)\, \widetilde\Sigma_\lambda({\cal M}^2)\,
 {k^2 \over k^2 + \widetilde\Sigma_\psi^2(k^2)}
 \Biggr]\ .
\end{eqnarray}
In Eqs.~(\ref{squarkDgap}) and (\ref{squarkXgap}) we have dropped some
contributions from diagrams with higher powers of mass insertions, as well
as higher order contributions from diagonalizing the scalar propagator.
Supersymmetry ensures cancellation of the quadratic divergences in the
diagonal mass equation (\ref{squarkDgap}).  After that cancellation,
both equations can be fully linearized for momenta $p$ large compared to
$\widetilde\Sigma$.

   Then up to mass-suppressed corrections, $A_\phi(p^2)$ is fixed by
Eqn.~(\ref{squarkDgap}), while
$A_\lambda(p^2) \approx (\mu^2/p^2)^{\gamma_\lambda}$
and $A_\psi(p^2) \approx (\mu^2/p^2)^{\gamma_\psi}$, where
\begin{eqnarray} \label{anomdimSQCD}
 \gamma_\lambda\ & = &
 {g^2 \over 16\pi^2}\, ( \xi\, C_2^{\rm adj} +  2 N_f\, C_{\rm fund} )\ ,
\nonumber\\
 \gamma_\psi\ & = & {g^2 \over 16\pi^2}\, (\xi + 1)\, C_2^{\rm fund}.
\end{eqnarray}
Here $C_{\rm fund} = 1/2$, and $g^2$ and $\xi$ are the fixed point
values described above.
Since the critical couplings depend on the wave function factors,
the gluino critical coupling will differ from its value in the matter-free
theory, even though in both cases gluon emission and reabsorption is the
only leading contribution to the gap equation.

   We focus initially on the quark and gluino equations.  As described
for the matterless theory, substituting the wave function factors of
Eq.~(\ref{anomdimSQCD}) into the gap equations, and setting the mass anomalous
dimensions to unity, leads to critical couplings independent of $\xi$:
\begin{eqnarray} \label{critcouplSQCD}
 \alpha_{{\rm cr}, \lambda} & = &
 { \pi\over 3\, C_2^{\rm adj} - 2 N_f\, C_{\rm fund} }
 \ =\ { \pi\over 3\, N_c - N_f }\ ,
\nonumber\\
 \alpha_{{\rm cr}, \psi} & = & \qquad
 { \pi \over 2\, C_2^{\rm fund} } \qquad\ \,
 \ =\ { \pi\over N_c - 1/N_c }\ .
\end{eqnarray}
The choice of $N_f$ determines whether the infrared coupling $\alpha_*$
in Eq.~(\ref{fixedpoint}) achieves these critical values.
As $N_f$ is reduced from $3 N_c$, $\alpha_*$ increases from zero and
first exceeds the quark critical coupling,  $\alpha_{{\rm cr}, \psi}$,
when $N_f/N_c = ({9\over 4}) { 1 - (2/3) N_c^{-2} \over 1 - (3/4) N_c^{-2} }
\approx 2.25$.  For this value of $N_f / N_c$,
$\alpha_*$ does not yet reach the gluino critical coupling,
leaving quark condensation to play the primary role here.  Assuming validity
of the approximations above, we can now outline the phase structure of SQCD
as described by gap equations:
\\ \noindent
$\bullet$ For $N_f > 3 N_c$ there is no asymptotic freedom, in agreement
with the holomorphic description.
\\ \noindent
$\bullet$ For $2.25 N_c < N_f < 3 N_c$ there is an infrared fixed point
at which the two-loop $\beta$ function vanishes.  For $N_f$ close enough to
$3 N_c$, the fixed point coupling is small, keeping the coupling well below
the critical values of Eq.~(\ref{critcouplSQCD}).  For any $N_f$ above
$2.25 N_c$, the infrared coupling remains below both critical couplings,
leaving the quark and gluino symmetries unbroken.
Confinement does not set in and the theory remains in the nonabelian
Coulomb phase, in agreement with the holomorphic prediction.
\\ \noindent
$\bullet$ For $N_f < 2.25 N_c$, the infrared coupling (in the two-loop
approximation) exceeds the quark critical coupling.  If higher order effects
may be neglected, the gap equation then indicates that quarks develop a
dynamical mass which vanishes continuously at the transition.  The quarks
decouple from the $\beta$ function at this mass scale, eliminating the fixed
point and causing $g$ to run to the gluinos' (larger) critical coupling.
Gluino and scalar dynamical masses are then generated at essentially the
same scale as the quark mass.
Below these mass scales, decoupling allows $g$ to continue running to
confining values.  The global chiral and $U(1)_X$ symmetries are thus
predicted to break in this $N_f$ range.  The supersymmetric Ward identity
then requires that supersymmetry breaks as well.
\\
Holomorphic methods, by contrast, predict the nonabelian Coulomb phase to
remain down to $N_f = {3\over 2} N_c$, and allow preservation of the global
symmetries without confinement even down to $N_f = N_c + 1$.

   These patterns also control the scalar gap equation solutions, which
depend upon the fermion solutions.  The inhomogeneous scalar equations
do not admit vanishing solutions when the quark and gluino dynamical
masses are nonzero \cite{shamir}.  In that case, the scalar solutions
should be of comparable scales.  The connection between scalar and fermion
mass solutions can be understood from the global symmetries:  a nonzero
off-diagonal mass $\widetilde\Sigma_X^2(p^2)$ breaks both $U(1)_X$
and $SU(N_f) \times SU(N_f)$, and thus requires corresponding gluino and
quark masses.  The diagonal mass $\widetilde\Sigma_\phi^2(p^2)$ breaks no
chiral or $U(1)$ symmetries, but is not forced to vanish unless the quark
mass does and supersymmetry is unbroken.

   Checking that these symmetry breaking solutions of the gap equation
describe the ground state would require showing that the resulting vacuum
energy is lower than that of the chirally symmetric solution.
(For this to be the case the symmetric solution, as well as the broken
one, would need to break supersymmetry.)  For SQCD with matter,
suggestive calculations \cite{shamir,rothstein-shamir} do indicate lower
energy for the symmetry breaking vacuum, although these are gauge dependent
and neglect squark and/or gluino dynamical masses.

\section{Conclusion}

   In pure SQCD without matter fields, various approaches indicate that
the discrete global symmetry $Z_{2 N_c}$ breaks spontaneously to $Z_2$
through the formation of a gluino condensate.
The theory confines and supersymmetry presumably remains unbroken.
In the gap equation, the lowest order kernel, single gluon exchange,
provides an attractive force capable of condensate formation.
Whether the broken solution represents the ground state of the theory,
however, remains unclear.  Assuming this to be the case, the gap equation
and holomorphic effective action techniques lead to the same conclusion.

   For SQCD with matter multiplets, gluon exchange alone again gives the
leading contribution to the linearized gap equation governing the transition.
At this order, the gap equation (with gauge coupling governed by the two-loop
$\beta$ function) indicates that spontaneous breaking of the global symmetries
occurs for $N_f$ below a critical value near $2.25 N_c$.  
This transitional value corresponds to the infrared fixed point coupling
exceeding the critical strength necessary for quark condensation.  
If we instead determine the fixed point coupling from the three-loop $\beta$
function (in the DR scheme) \cite{SQCDbeta}, the corresponding critical
$N_f \approx 2.08 N_c$, only an 8\% shift in this renormalization scheme.
Examining the size of higher order terms in the kernel of the quark gap
equation, corresponding to the mass anomalous dimension at two loops,
would give another check on the validity of these approximations.
In nonsupersymmetric theories, such kernel corrections are of order 20\%
\cite{ALM,Nash}.  With these caveats, for $N_f$ just below the critical value
the quark mass is nonzero but small, justifying the linearization.

   At scales below their predicted mass the quarks decouple, eliminating the
infrared fixed point so that the coupling increases to the gluino critical
value at essentially the same scale.  The gluinos, too, thus condense at the
quark mass scale.  Squark masses are induced by the fermion masses, and are
also of the same scale.  The supersymmetric Ward identity then predicts that
supersymmetry also breaks in this regime.  An effective theory with Goldstone
bosons and Goldstinos emerges at low energies.
All this is in contrast with the picture emerging from holomorphic effective
action techniques, where the nonabelian Coulomb phase is argued to persist
down to $N_f = {3\over 2} N_c$ and the global symmetries to remain unbroken
(at the origin of moduli space) down to $N_f = N_c + 1$.

   Of course, as already stressed, gap equation solutions in general have
not been shown to correspond to ground states.  (Supersymmetric constraints
on ground state energies may be important in this connection.)
Settling the question involves still-unresolved issues of gauge invariance
and interpretation of the effective potential.
Furthermore, truncating kernels at one loop, and the $\beta$ function at two
loops, seems quantitatively less reliable than in nonsupersymmetric theories.
The kernel truncation, together with the use of component notation
(Wess-Zumino gauge), may also in effect explicitly break supersymmetry.
The latter possibility can be checked by analyzing a set of gap equations
in the manifestly supersymmetric superspace formalism \cite{ClarkLove}.
A recent study of this problem \cite{KaiserSelipsky} indicates that
symmetry-breaking solutions could exist, although infrared divergences
in the formalism have so far obscured the analysis.

   The possible discrepancy between gap equation analyses and results based
on holomorphic effective action techniques and index theorems is noteworthy
and invites further study.  Efforts to reconcile them may help deepen our
insight into the behavior of strongly coupled gauge theories.

\begin{flushleft} {\Large\bf Acknowledgments} \end{flushleft}

   The authors are grateful to Nick Evans, Stephen Hsu and Myckola Schwetz
for their assistance in the early stages of this work.  We thank them and
Andrew Cohen, Noriaki Kitazawa, Aneesh Manohar, Gregory Moore, Pierre Ramond
and John Terning for helpful discussions.  This work was supported in part
under United States Department of Energy DOE-HEP Grant DE-FG02-92ER-4074.
A.N. also acknowledges the support of Schweizerischer Nationalfonds.


\end{document}